# Residual Connection Networks in Medical Image Processing: Exploration of ResUnet++ Model Driven by Human Computer Interaction

Peixin Dai[1,3], Jingsi Zhang[2,4], Zhitao Shu[3,5]

[1]School of Electronic and Information Engineering, Lanzhou Jiaotong University, China
[2]Khoury College of Computer Sciences, Northeastern University, Boston, MA, USA
[3]Vanderbilt University, Nashville, TN, USA

[3]313290632@qq.com
[4]zhang.jings@northeastern.edu
[5]zhitao.shu@vanderbilt.edu

**Abstract.** Accurate identification and localisation of brain tumours from medical images remain challenging due to tumour variability and structural complexity. Convolutional Neural Networks (CNNs), particularly ResNet and Unet, have made significant progress in medical image processing, offering robust capabilities for image segmentation. However, limited research has explored their integration with human-computer interaction (HCI) to enhance usability, interpretability, and clinical applicability. This paper introduces ResUnet++, an advanced hybrid model combining ResNet and Unet++, designed to improve tumour detection and localisation while fostering seamless interaction between clinicians and medical imaging systems. ResUnet++ integrates residual blocks in both the downsampling and upsampling phases, ensuring critical image features are preserved. By incorporating HCI principles, the model provides intuitive, real-time feedback, enabling clinicians to visualise and interact with tumour localisation results effectively. This fosters informed decision-making and supports workflow efficiency in clinical settings. We evaluated ResUnet++ on the LGG Segmentation Dataset, achieving a Jaccard Loss of 98.17%. The results demonstrate its strong segmentation performance and potential for real-world applications. By bridging advanced medical imaging techniques with HCI, ResUnet++ offers a foundation for developing interactive diagnostic tools, improving clinician trust, decision accuracy, and patient outcomes, and advancing the integration of AI in healthcare workflows.

## 1. Introduction

The brain, as the core of the central nervous system, plays a vital role in controlling complex bodily functions. Brain tumours, however, pose significant threats to these functions, potentially leading to neurological impairments or fatal outcomes. Therefore, the accurate detection and delineation of tumour regions are crucial for prompt treatment planning and effective tumour classification.

Traditional brain tumor detection methods, including clinical symptom assessment, spinal puncture, and imaging, have limitations. Clinical symptoms can be ambiguous, and spinal taps carry risks, particularly with large tumors or high intracranial pressure. Imaging, especially MRI, is the most effective method for tumor detection, providing detailed insights into size and location. However, diagnostic errors can still occur, particularly in difficult-to-reach areas or irregular shapes. Deep learning techniques, particularly UNet and nn-UNet, have significantly improved medical image processing. These models excel in detecting and segmenting brain tumors in MRI images, outperforming traditional methods. Despite their success, challenges remain in enhancing model generalization and accuracy for complex cases.

This study explores the use of deep learning models, particularly combining ResNet and UNet, to enhance tumour detection and segmentation in MRI brain images. We introduce ResUnet++, a hybrid model with residual connections to preserve critical features during both downsampling and upsampling, improving segmentation performance. The model also incorporates data pre-processing and augmentation techniques for better generalisation. Additionally, human-computer interaction (HCI) principles are integrated to improve usability, allowing clinicians to engage with real-time visualisation and feedback. This research aims to enhance brain tumour detection and localisation, improving diagnostic accuracy and AI integration in healthcare.

## 2. Literature survey

Convolutional Neural Networks (CNNs) have seen significant advancements in medical image processing, with deep learning models like Unet Networks leading the way. In 2015, Kaiming He and Xiangyu Zhang introduced ResNet[1], which overcame deep network training challenges with residual learning, successfully training a 152-layer network that excelled in ILSVRC 2015. ResNet's design allows for direct information flow through shortcut connections, mitigating gradient vanishing and network degradation issues.

In 2016, Olaf Ronneberger, Philipp Fischer, and Thomas Brox introduced Unet, a network known for capturing contextual information through hopping connections, enhancing segmentation accuracy[2]. Unet's success was proven by winning the 2015 ISBI Cell Tracking Challenge. In 2018, Qiangguo Jin and colleagues advanced the field with RA-UNet[3], a 3D segmentation model integrating recurrent attention and residual learning, achieving high Dice scores in ISBI and MICCAI challenges. Also in 2018, Ozan Oktay and colleagues introduced Attention U-Net[4], which incorporated attention mechanisms to improve sensitivity by focusing on relevant target structures and suppressing irrelevant areas.

In 2020, Huimin Huang, Lanfen Lin, and colleagues introduced UNet 3+[5], which expanded upon UNet++ by introducing full-scale jump connectivity and deep supervision. This redesign of decoder-encoder interconnections and internal decoder links resulted in significant improvements over UNet and

UNet++ on liver and spleen datasets from the ISBI LiTS 2017 challenge, outperforming five contemporary models including PSPNet[6] , DeepLabV2[7] , DeepLabV3[8] , DeepLabV3+[9] , and Attention UNet.

In 2023, Asbjørn Munk, Ao Ma, and Mads Nielsen proposed an unsupervised domain adaptive framework based on U-Net and Margin Disparity Discrepancy (MDD), improving segmentation across 11 of 12 dataset combinations[11]. In 2024, Jiacheng Ruan and Suncheng Xiang introduced VM-UNet, a U-Net architecture with visual state-space (VSS) blocks, offering superior performance on ISIC17, ISIC18, and Synapse datasets, with significantly higher IoU, DSC, and Acc metrics than traditional U-Net[12].

**3. Segmentation Model**

3.1 Data processing

In order to be model training effect is more effective and facilitate the convenience of the subsequent operation, we carry out the relevant processing of the data set, the relevant data processing techniques are: we enhance model generalization by randomly shuffling the image order at the start of each training cycle, stabilize data through standardization by subtracting the mean and dividing by the standard deviation, normalize pixel values to a range of 0 to 1 for binary classification, and ensure uniform input by adjusting image sizes.

3.2 Resunet++

We propose a novel convolutional neural network model named ResUnet++, which combines the advantages of residual networks (ResNet) and U-Net architectures to improve the performance of medical image segmentation tasks. ResUnet++ facilitates the flow of information in the deep network by introducing residual blocks, while utilizing the encoder-decoder structure of U-Net to precisely locate the target regions in the image.

Compared to the traditional ResUnet model, ResUnet++ further optimizes the network architecture, enhancing feature extraction and reconstruction capabilities. It excels in handling complex image segmentation tasks, particularly in medical imaging, where it demonstrates superior precision and robustness when dealing with challenging tumor regions. ResUnet++ incorporates the ASPP module (Atrous Spatial Pyramid Pooling), which captures multi-scale information through dilated convolutions, enhancing the model's ability to extract features at different scales. This allows the network to better adapt to the complex structures of medical images. The attention mechanism helps the model automatically focus on important areas of the image during segmentation, suppressing irrelevant regions, thereby improving segmentation accuracy, especially in blurry or noisy areas.

Through these enhancements, ResUnet++ offers a more accurate and efficient solution for medical image segmentation tasks, particularly in tumor detection and localization.

(1) Resunet++ downsampling section

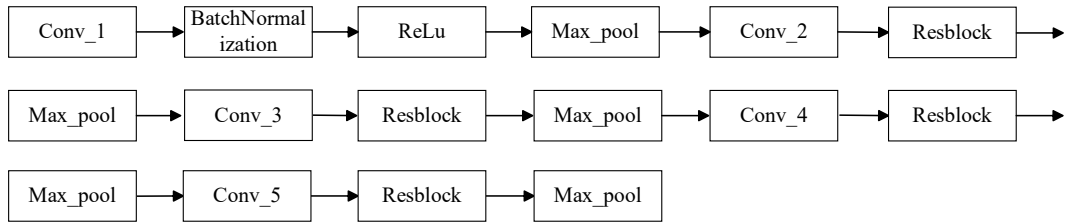

Fig. 1. Resunet++ downsampling section

(2) Resunet++ upsampling section

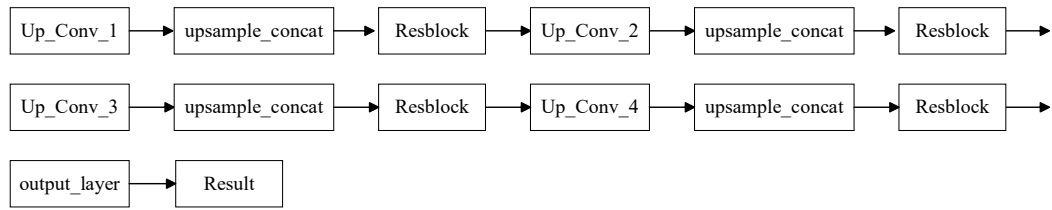

Fig. 2. Resunet++ upsampling section

The input image is 256×256×3. The first convolutional layer uses a 3×3×16 kernel (learning 16 features) with BatchNormalization and ReLU activation. A Maxpool layer halves the size. The second convolutional layer also uses a 3×3 kernel, learning 32 features, followed by a Resblock for residual connections, then another Maxpool. The third, fourth, and fifth layers work similarly, learning 64, 128, and 256 features, resulting in a final size of 16×16×256.

For downsampling, a 2×2 upsampling layer doubles the image size, with residual connections via a Resblock. The upsampling layers mirror the downsampling process, leading to a final output size of 256×256×16. A 1×1×1 convolution kernel reduces the output to a single channel.

The segmentation model uses EarlyStopping, Model Checkpointing, and Learning Rate Reduction. We replace traditional loss functions with Jaccard Loss, which calculates the ratio of intersection to union between predicted and true regions. This loss function handles class imbalance well and improves segmentation accuracy. The hyperparameters include the Adam optimizer with a learning rate of 0.05, epsilon of 0.1, and 60 epochs of training.

In ResUnet++, the ASPP module (Atrous Spatial Pyramid Pooling) captures multi-scale information using dilated convolutions, enhancing the model's ability to detect tumor regions. Additionally, the attention mechanism helps the model focus on relevant features and suppress irrelevant ones, improving segmentation performance, particularly in noisy or blurred medical images. This combination enhances the model's robustness in medical image segmentation tasks.

3.2 NAdam optimizer and learning rate

NAdam optimizer is an improved version of Adam, combining the advantages of Nesterov Accelerated Gradient (NAG) and Adam. It incorporates momentum to update parameters more efficiently, accelerating the training process. NAdam computes adaptive learning rates by estimating the first (mean) and second (variance) moments of gradients. Compared to Adam, NAdam uses Nesterov's momentum, which helps optimize training stability and speed. In this paper, we chose NAdam for its ability to handle gradient changes effectively, ensuring more stable updates. We also implemented learning rate decay over 10 cycles to improve optimization precision.

**4 Experimental results**

4.1 Introduction to the datasets

The dataset in this paper is from Mateusz Buda, Ashirbani Saha, and Maciej A. Mazurowski's study, "Association of genomic subtypes of lower-grade gliomas with shape features automatically extracted by a deep learning algorithm" (2019). It includes MRI brain images and corresponding mask files marking tumor regions, with black areas indicating no tumor. The dataset is used for training, validation, and testing the model.

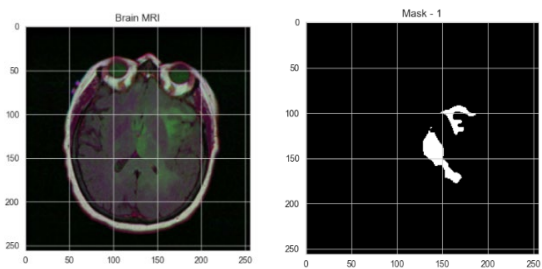

Fig. 3. A sample dataset

4.2 Experimental results of ResUnet image segmentation algorithm

　　Our IoU index reaches 98.17%, indicating a good model fit.

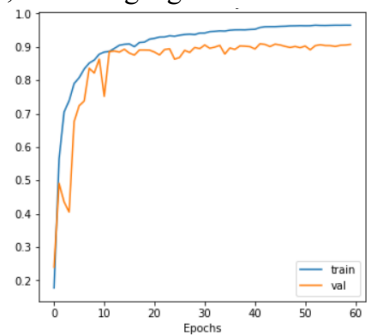

Fig. 4. The change of accuary for resunet++

4.3 ResUnet image segmentation algorithm prediction results

　　For the problem addressed in this paper, we first use a classification model to detect the presence of a tumor in the image, and if it exists, a segmentation model is used to accurately divide the tumor region to predict the result as follows

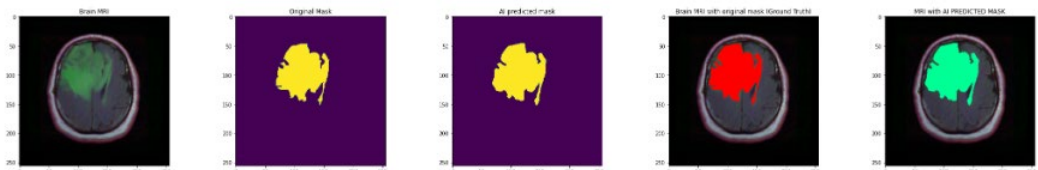

Fig. 5. Prediction results

　　In figure10, Image 1 is the original brain MRI image, image 2 is the corresponding mask image of the original this brain, the yellow area is the tumor area, image 3 is the tumor area predicted with the model, image 4, 5 are the positions of the original tumor area and the predicted tumor area in the brain

MRI image, corresponding to the red and green areas, respectively. It can be seen that the prediction results basically match.

## 5. Conclusion

This study aims to achieve the precise identification and localization of brain tumors using the ResUnet++ segmentation algorithm. In the tumor region segmentation task, we combine the residual connections of ResNet with the basic framework of U-Net to propose a new ResUnet++ model. The model incorporates the ASPP module and attention mechanism, further enhancing its ability to capture multi-scale information and focus on key regions. The final model uses the Jaccard loss function, achieving a Jaccard index of 98.17%. After 60 training epochs, the training set loss rate is 0.0764, and the validation set loss rate is 0.1718, demonstrating excellent performance.

The model effectively identifies brain tumor images and segments tumor regions, demonstrating its potential in image recognition despite some overfitting and accuracy fluctuations. Limitations include dataset quality and performance on rare tumor subtypes or image artifacts. Future research could explore incorporating attention mechanisms, lightweight techniques, or multimodal fusion with other models. Additionally, using data augmentation and Generative Adversarial Networks (GANs) to generate high-quality simulated data could enhance the model's generalization and performance.